\documentclass[a4paper]{article}

\usepackage{multirow}
\usepackage{INTERSPEECH2019}
\usepackage{xcolor}
\usepackage{multirow}

\newcommand{\ms}{\mathbf{S}}

\newcommand{\newpara}[1]{\vspace{5pt}\noindent\textbf{#1}}

\title{Metric Learning for Keyword Spotting}
\name{Jaesung Huh*, Minjae Lee*, Heesoo Heo*\thanks{\hspace{-12pt}* These authors contributed equally to this work.}, Seongkyu Mun, Joon Son Chung}
\address{Naver Corporation, South Korea}
\email{\{jaesung.huh,mjlee.0328,heesoo.heo,sk.moon,joonson.chung\}@navercorp.com}

\begin{document}

\maketitle

\begin{abstract}
The goal of this work is to train effective representations for keyword spotting via metric learning. 
Most existing works address keyword spotting as a closed-set classification problem, where both target and non-target keywords are predefined.
Therefore, prevailing classifier-based keyword spotting systems perform poorly on non-target sounds which are unseen during the training stage, causing high false alarm rates in real-world scenarios. 
In reality, keyword spotting is a detection problem where predefined target keywords are detected from a variety of unknown sounds.
This shares many similarities to metric learning problems in that the unseen and unknown non-target sounds must be clearly differentiated from the target keywords. 
However, a key difference is that the target keywords are known and predefined.
To this end, we propose a new method based on metric learning that maximises the distance between target and non-target keywords, but also learns per-class weights for target keywords as in classification objectives. 
Experiments on the Google Speech Commands dataset show that our method significantly reduces false alarms to unseen non-target keywords, while maintaining the overall classification accuracy. 
\end{abstract}


\section{Introduction}
\label{sec:intro}

 Keyword Spotting (KWS) is the task of detecting small set of predefined speech signals such as wakeup words of various mobile devices (e.g. ``OK Google", ``Hey Siri" and ``Alexa") or frequently-used short commands. CNN based architectures have recently achieved state-of-the-art performance in this field \cite{sainath2015convolutional,tang2018deep, choi2019temporal, zhang2017hello, mittermaier2020small, zeng2019effective, coucke2019efficient}, which are primarily based on classifiers that distinguish between each of the target keywords of interest and non-target sounds such as general speech and noises. Although the non-target sounds can be extremely diverse, only a limited number of non-target classes have been used in previous works, resulting in poor generalisation to real-world scenarios. Traditional literature have attempted to reduce false alarms via post-processing \cite{lubovs2009keyword}, but such methods have not been used in conjunction with deep learning approaches.
 
 Keyword spotting in the real-world is similar to a detection problem rather than that of classification, where the predefined keywords are to be spotted from a range of unknown sounds.
 However, many previous works have considered the non-target sounds as a single class~\cite{tang2018deep, choi2019temporal, zhang2017hello, chen2014small, xu2020depthwise}.
 In order to bridge this gap, we take inspirations from metric learning approaches that learn discriminative embeddings that can be used to accept or reject target spoken terms. Although the types of keywords are pre-defined unlike in speaker verification, the problem is naturally closer to that of verification than of classification. 
 
 Metric learning objectives such as the contrastive loss \cite{koch2015siamese, chopra2005learning} and triplet loss \cite{hoffer2015deep, chechik2010large} have been used widely in both face recognition \cite{parkhi2015deep, schroff2015facenet} and speaker verification \cite{nagrani2017voxceleb, zhang2017end}. The metric learning methods map the input signals into an embedding space, enlarging inter-class variance and reducing intra-class variance. 
 
 Recent metric learning techniques have been introduced to overcome the weaknesses of contrastive and triplet losses, namely the difficulty of pair selection. \cite{wan2018generalized} and \cite{snell2017prototypical} have proposed training frameworks that do not require careful pair-selection by using multiple positives and negatives during training, and this has shown to improve the performance in speaker verification~\cite{wan2018generalized,chung2020defence, heo2019end}. 
 A related approach in keyword spotting is \cite{gundougdu2017joint}, that uses Siamese neural network~\cite{chopra2005learning} to train frame-wise embeddings that is subsequently used for speech recognition using Dynamic Time Warping.  
 
 In this paper, we present a number of novel strategies for keyword spotting inspired from metric learning methods. While keeping the network architecture constant, we change the loss functions from the classification loss to a range of metric learning objectives. 
 We train the embeddings to minimise intra-class distance for target classes, but not between non-target embeddings, which mimics the real-world problem of keyword spotting. This is in contrast to classification approaches that enforce similarity between potentially infinite types of non-target sounds. 
 On the popular Google Speech Command Dataset~\cite{warden2018speech}, we show that our proposed methods outperform the classification baseline for the detection task, while maintaining accuracy on the classification task.
 
 Our contributions can be summarized as follows.
 (1) In contrast to previous works in keyword spotting, we reformulate the problem as one of {\em detection}.
 (2) We propose methods inspired by  metric learning that can boost non-target keyword accuracy by a significant margin. This approach is evaluated on the experimental setting that mimics the real-world scenario where non-target sounds are not seen during training.
 (3) We propose a new inference method using support vector machine (SVM) that is applicable to networks trained with metric learning in keyword spotting.
 
 In the rest of this paper, we discuss existing metric learning approaches in Section \ref{sec:metric} with our new loss function for keyword spotting as well, experimental setups at Section \ref{sec:exp} and results and conclusion in Section \ref{sec:results} and \ref{sec:conclusion}, respectively.

\section{Metric learning framework}
\label{sec:metric}

In this section, we describe the existing loss functions used in metric learning, and propose modifications to boost non-target accuracy while maintaining the overall classification accuracy.

\subsection{Loss functions}
\label{subsec:loss}
Here, we explain the triplet loss and prototypical losses, which are used widely  in metric learning.

\newpara{Triplet loss.}
Triplet loss is one of the most common ranking loss function. Minimizing this loss function decreases the distance between the embeddings from the same class and increases the distance between ones from different classes at the same time. Let $f(\mathbf{x};\mathbf{w}) \in \mathbb{R}^D$ be a neural network that maps the input to corresponding embedding. $\mathbf{x}_{i}$, $\mathbf{x}^\prime_{i}$ are input samples from same class $i$, and $\mathbf{x}_{j}$ as a sample from different class $j \neq i$. $\lVert \mathbf{x} - \mathbf{y} \rVert$ is the pairwise-distance between $\mathbf{x}$ and $\mathbf{y}$. For each triplet $P_{i,j} = (\mathbf{x}_{i}, \mathbf{x}^\prime_{i}, \mathbf{x}_{j})$, triplet loss $L$ is minimized per batch level
\begin{equation}
\label{eq:triplet}
\begin{aligned}
    L {} & =  \sum_{i} L_{T} (P_{i,j};\mathbf{w}) \\
         & = {\sum_{i} \max{( 0, \lVert f(\mathbf{x}_{i}) - f(\mathbf{x}^\prime_{i}) \rVert - \lVert f(\mathbf{x}_{i}) - f(\mathbf{x}_{j}) \rVert + \alpha )}},
\end{aligned}
\end{equation}
where $\alpha$ is a constant margin. (e.g. $\alpha = 1$)

\newpara{Prototypical networks.}
Prototypical networks have been proposed by~\cite{snell2017prototypical} to learn a metric space in which open classification can be performed. This is done by computing distance from an embedding to the prototype representations of each class, which is relevant to our work.

In our implementation, we use an angular variant of the prototypical loss~\cite{chung2020defence} which replaces the distance metric in the original paper with that from~\cite{wan2018generalized}. Here, each mini-batch contains $N \times M$ input features, from $N$ different classes with $M$ utterances per class. Assuming $\mathbf{e}_{j,M}$ as the query of class $j$ in each batch, angular prototypical loss uses a cosine-based similarity metric with learnable scale and bias. It uses angular loss for stable and robust training instead of L2 distance. A scalable cosine-similarity between embedding of $M$-th utterance of class $j$ and centroid of $k$-th class $\mathbf{c}_{k}$ defined as:

\begin{equation}
\ms_{j,k} = 
w\cdot \cos(\mathbf{e}_{j,M}, \mathbf{c}_{k})+b, 
\label{eqn:angleproto_dist1}
\end{equation}
\begin{equation}
\mathbf{c}_{k} = \frac{1}{M-1}\sum_{i=1}^{M-1}\mathbf{e}_{k,i},
\label{eqn:angleproto_dist2}
\end{equation}

\noindent where $w$ and $b$ are learnable parameters with constraint $w > 0$. For each batch, the objective of angular prototypical loss is maximizing the similarities between embeddings and the corresponding centroid of the same class, while minimizing the similarities between embeddings and the centroids from different classes. To achieve this objective, the loss function is defined as:

\begin{equation}
L = -\frac{1}{N} \sum_{j=1}^{N} \log
\frac{e^{\ms_{j,j}}}
{\sum_{k=1}^N e^{\ms_{j,k}}}.
\label{eqn:proto_loss}
\end{equation}

\subsection{Pair selection strategy}
\label{sec:pairs}
Here, we introduce the strategies for training network using the loss functions in Section~\ref{subsec:loss}. 
Specifically, we address how positive and negative pairs should be selected to effectively distinguish `target' keywords from unknown `non-target' sounds.

\newpara{Metric learning with an unknown cluster.}
This is the baseline metric learning approach using the loss functions in  Section \ref{subsec:loss}. In this approach, target keywords, as well as non-target keywords, are clustered to an anchor or a centroid in the embedding space using the triplet or prototypical losses. In other words, the network is trained to learn features that have small intra-class and large inter-class variance for both target and non-target keywords. Drawback of this approach is treating non-target keywords as one class similar to target keywords, although they show much higher variance compared to each of target keywords.

 Since we do not train the network with a classification objective, we propose a new inference method to make decisions. Having trained the network, we extract embedding vectors from the whole training data and compute centroids of each class by averaging them. During the testing stage, we compute the embedding for each test sample and calculate the similarities with all centroids to decide which class to the sample belongs to.
 

 This can be also easily implemented in the real world scenarios. The centroids that correspond to each class are pre-calculated and considered as model parameters. When a target input is given, an embedding vector can be extracted with trained model. 
 By computing distances between the extracted embedding vector and each centroids, we can classify given input data. 
 
\newpara{Metric learning without an unknown cluster.}
 The range of non-target keywords is much larger than each of the target keywords, because the former contains all other sounds and speeches except the latter. However, in the conventional approach, all non-target sounds are assigned to one single class, and the network is trained to converge the embeddings to a single point.
 Gathering non-target sounds to a point in the feature space does not reflect the variance of the non-targets in real-world conditions. It is challenging to generalize the model trained using the limited types of non-target words to unseen words. Therefore, we propose a modification to {\em not} cluster the non-target keywords to one point during training. Objective functions in Equation \ref{eq:triplet} and \ref{eqn:proto_loss} are respectively modified as follows:
 
\begin{align}
\label{eqn:tripletmethod2}L = \sum_{i \in \{ target \} } L_{T} (P_{i,j}; \mathbf{w}), \\
\label{eqn:anglemethod2}L = -\frac{1}{N} \sum_{j \in \{ target \}}^{} \log
\frac{e^{\ms_{j,j}}}
{\sum_{k=1}^N e^{\ms_{j,k}}}.
\end{align}

 For discriminating the non-target words, we have to put an additional step since it is unable to use a centroid of non-target keywords. After training embedding extractor using metric learning, we train one-vs-rest radial basis function (RBF) kernel SVM with embeddings of training dataset \cite{platt1999probabilistic, chang2011libsvm}. Then we utilize this SVM to make final decision in the test set.
 
\subsection{Prototypical networks with fixed target classes}
\label{subsec:modifiedangularprototypicalloss}
Here, we propose a modified prototypical loss for target keyword spotting. 
In the original framework of prototypical networks, the centroids are computed during inference in a few-shot learning setting.
However, we can take advantage of the fact that the target keywords are fixed, unlike previous applications of prototypical networks such as face and speaker verification.
Therefore for known keywords, we replace the centroids in each class $\mathbf{c}_{k}$ that are computed on-the-fly with per-class weights $\mathbf{W}_{k}$ that are learnt.
Our empirical experiments show that the classifier-based keyword spotting systems have higher accuracy on target keywords, while false alarm rate on non-target keywords is lower with the metric learning-based systems. 
The proposed method incorporates advantages of both methods by detecting known keywords using the trained per-class weights, while being able to reject non-targets in a metric learning-like fashion.
We refer to the proposed method as angular prototypical with fixed classes (AP-FC).

We can define $\ms_{j,i,k}$ as a scaled cosine similarity between the $i$-th embedding of class $j$ and learnable parameter (anchor) $\textbf{W}_k$ for the $k$-th target keywords instead of the centroids.

\begin{equation}
\ms_{j,i,k} = w\cdot \cos(\textbf{e}_{j,i},\textbf{W}_k)+b \text{,} \quad k \in \{ target \}.
\label{eqn:KWSAPdistance}
\end{equation}

It is important to note that $\textbf{W}_k$ are the only learnable parameters for target keywords, and these parameters have a role of the output layer in the classifier. 
Then the loss function for AP-FC is defined as:

\begin{equation}
L = -\frac{1}{N-1} \sum_{k=1}^{N-1} \log
\frac{e^{\ms_{k,1,k}}}
{\sum_{j=1}^{N-1} e^{\ms_{j,1,k}} + \sum_{i=1}^{N^{\prime}} e^{\ms_{N,i,k}} },
\label{eqn:KWS-APloss}
\end{equation}

\noindent where $N^{\prime}$ is the number of samples of non-target keyword contained in one mini-batch. We assume a class index of non-target keywords is $N$ in the Equation \ref{eqn:KWS-APloss}.
We expect that the learnable parameters $\textbf{W}_k$ for all $k \in \{target\}$ are trained to have a role of the centroid of each target keyword. 
Note that one mini-batch contains one sample for each target keyword and multiple samples of the non-target keyword to adjust the balance of target and non-target classes. In particular, we set the value of $N^{\prime}$ to six in our experiments.

\section{Experiments}
\label{sec:exp}
In this section, we explain the experimental setup. The input and the layer configurations are consistent in all of our experiments. Therefore, we focus on comparing the performance of trained networks with a number of learning objectives.

\subsection{Dataset}
\label{subsec:dataset}
 Google Speech Commands v0.01 \cite{warden2018speech} is a popular keyword spotting dataset which contains 64,727 utterances of 30 different spoken terms from 1,881 speakers. As any other papers using this dataset, we select 10 target spoken terms (``Yes", ``No", ``Up", ``Down", ``Left", ``Right", ``On", ``Off", ``Stop" and ``Go") and add ``silence"(no speech signal) and ``unknown" class, which are rest of 20 non-target keywords. Spoken words in the ``unknown" class are the non-target signals in our case.
 
 We change the train-test split for unknown keywords to reflect the likely real-world usage. In the original dataset, the words contained in ``unknown" class of train and test set overlap -- the same list of ``unknown" words that are used during training is also used for testing. In reality, ``unseen-unknown" words, the non-target keywords not seen by the model during training, are seen more often by a KWS model compared to the ``seen-unknown" words. Therefore, it is important to correctly classify unseen-unknown words as unknown to reduce False Alarm rate. 
 
 To reflect this, while leaving the train-test split for the target keywords the same as in the original dataset, we split the 20 ``unknown" keywords to half, and use 10 words for training and the other 10 words for testing. See Table~\ref{tab:dataset} for the exact split. The ``unknown" keywords in train set are ``seen-unknown" words, while ``unknown" words used in test set are ``unseen-unknown". 
 Unseen-unknown utterances in the former training and test set are used in the new test set, since the overlap between unseen-unknown keywords and the former test set is too small ($<100$).
 We report results on two test lists -- one with known to unknown ratio of 11:1 which is the ratio in the original dataset, and one with 1:1 which more closely reflect real-world conditions. The results on these are discussed in Section \ref{sec:results}.
 
\begin{table}[t]
\caption{Dataset configuration. While maintaining the target keywords utterances in both train and test set of Google Speech Commands v0.01 \cite{warden2018speech}, 20 non-target keywords are divided into two categories, one for train set and another for test set.}
\label{tab:dataset}
\centering
  \renewcommand{\arraystretch}{1.2}
  \vspace{10pt}
\begin{tabular}{|c|c|c|}
\hline
\textbf{Dataset} & \textbf{Target keywords} & \textbf{Non-target keywords} \\ \hline
Train set   & \multirow{2}{*}{\begin{tabular}[c]{@{}c@{}}'Yes', 'No', \\ 'Up', 'Down', \\ 'Left', 'Right', \\ 'On', 'Off', \\ 'Stop', 'Go' \\ 'silence' \end{tabular}} & \begin{tabular}[c]{@{}c@{}}'Zero', 'One', \\ 'Two', 'Three', \\ 'Four', 'Five',\\ 'Six', 'Seven', \\ 'Eight', 'Nine'\end{tabular}    \\ \cline{1-1} \cline{3-3} 
Test set         &                                                                                                                                            & \begin{tabular}[c]{@{}c@{}}'Bed', 'Bird', \\ 'Cat', 'Dog', \\ 'Happy', 'House', \\ 'Marvin', 'Sheila', \\ 'Tree', 'Wow'\end{tabular} \\ \hline
\end{tabular}
\end{table}

\subsection{Feature Extraction and data augmentation}
\label{subsec:feature}
 We use 40 dimensional Mel-Frequency Cepstrum Coefficient (MFCC) of the original 16kHz waveform as an input, with 40ms frame length and 20ms hop size. All of the wav files are chunked to one second.
 
 For data augmentation, at each epoch,  20\% of the input features are randomly selected and shifted by $t \sim U[-10, 10]$ frames (i.e. -200ms $\sim$ 200ms) along time axis and zero-padded to match the size of other input features.
 
\subsection{Model Architecture}
\label{subsec:model}
 For the baseline network, we utilize the \texttt{res15} architecture proposed by \cite{tang2018deep}. It uses residual connections \cite{he2016deep} and dilated convolutions \cite{yu2015multi} to increase the receptive field of the network. Details of the architecture are shown in Table \ref{tab:architecture}. Each residual block contains two 3 $\times$ 3 convolution blocks with skip connection. All of the convolution blocks are followed by batch normalization \cite{ioffe2015batch} and ReLU function. Embedding vector size $\it{D}$ is fixed to a certain number (e.g. 32) in all of our experiments.
 
\begin{table}[t]
\caption{Backbone architecture of our network proposed by \cite{tang2018deep}. $D$ is the size of embedding vector.}
\label{tab:architecture}
\footnotesize
\centering
  \renewcommand{\arraystretch}{1.2}
  \begin{tabular}{| c | c | c|  c | c |c|}
    \hline
    \multirow{2}{*}{\bf Layer} 
    & \multicolumn{2} {c|} {\bf Width} & \multicolumn{2} {c|} {\bf Dilation} & \multirow{2}{*}{\bf Filters / Nodes} \\ \cline{2-5} 
    & time & freq & time & freq & \\ \hline
    cnn 1 &3&3&1&1&45\\
    res $\times$ 6 &3&3&$2^{\lfloor \frac{i}{3} \rfloor}$&$2^{\lfloor \frac{i}{3} \rfloor}$&45\\
    cnn 2 &3&3&16&16&45\\
    avg-pool &-&-&-&-&45\\
    fc 1 &-&-&-&-&$D$\\
    \hline
  \end{tabular}
 \normalsize
\end{table}


\begin{table*}[th]
\caption{Results of overall experiments. Baseline: network trained with cross entropy loss. AP: angular prototypical loss. All numbers are in percent. $\dag$ Known to unknown ratio of 11:1 matching the original test set; $\ddag$ known to unknown ratio of 1:1 which is more likely to reflect real world usage in keyword spotting.}
  \label{tab:result}
  \centering
\begin{tabular}{c c | c c c c  | c c}
\toprule
 &  & \multicolumn{4}{c|}{Classification metrics} & \multicolumn{2}{c}{Detection metrics} \\
Loss         & Back-end    & Total acc.$\dag$ & Total acc.$\ddag$ & Target acc. & Non-tgt acc. 
& ~~~AUC~~~ & ~~~mAP~~~ \\
\midrule
\multicolumn{2}{c|}{Baseline}          &  91.83  & 73.81                    & \textbf{95.52}    & 52.11                     &       95.63           & 94.84                        \\
\midrule
Triplet                  & Centroid  & 90.61 & 76.59               & 93.45              & 59.47                    &             \textbf{98.66}       &               92.80      \\
AP                       & Centroid & 92.46  & 77.80                & 95.40             & 60.22                    &     95.82                &      95.19               \\
\midrule
Triplet                  & SVM     & 90.39   & 78.96                & 92.68             & 65.24                      &           97.17     & 92.05                         \\
AP                       & SVM     & 91.47 & 76.40                & 94.49           & 58.30                        &             97.68     & 94.18                     \\
\midrule
AP-FC                   & SVM     & \textbf{92.52}   & \textbf{83.82}      & 94.26            & \textbf{73.37}             &             97.17    & \textbf{95.42} \\ \bottomrule                                                             
\end{tabular}
\end{table*}


 \subsection {Training}
 \label{subsec:traininfer}
 \vspace{3pt}\noindent
 Our networks are trained with PyTorch~\cite{paszke2017automatic}. Each model is trained for 150 epochs, using the Adam optimizer \cite{kingma2014adam} with momentum of 0.9. Initial learning rate is set to 0.001 and reduced by 0.1$\times$ if the validation accuracy does not improve for 10 epochs. We use a weight decay of $1e-5$ to alleviate overfitting. Hard negative mining is conducted for models trained with the triplet loss, which is crucial for improving performance. 
 For the prototypical loss, each batch has $M=6$ input features per $N=12$ classes. All of the methods are evaluated for five times and we report average of overall performance. Early stopping \cite{prechelt1998early} is employed using validation dataset.
 
 We additionally train the network as 12-way classification network on the original train-test split to validate the our implementation. We confirm the overall accuracy very close to the original work (95.5\% classification accuracy) and train the networks with the same implementation on our new split to fairly compare with our proposed experiments.
 
\subsection{Evaluation}
The primary metric in our experiments is classification accuracy, the proportion of correct decisions out of total number of samples. \textbf{Target Acc} is the classification accuracy of 11 target keywords except for the ``unknown" class. Also, we report \textbf{Non-target Acc} which refers the accuracy of the unknown keywords -- {\em i.e.} this is the probability of classifying an unknown utterance as unknown. 
We report two total accuracies, as described in Section~\ref{subsec:dataset}. \textbf{Total acc}$\dag$ is the accuracy which test set has target to unknown ratio of 11:1 which the ratio is same as the original dataset but with different non-target sounds. \textbf{Total Acc}$\ddag$ is overall accuracy to the test set where known to unknown ratio of 1:1 which is more likely to reflect real world usage in keyword spotting. 

We also report two metrics that are commonly used in detection, area under the curve (AUC) and mean average precision (mAP). All of the experiments were conducted five times and average performance are reported in here. 

We also report the Receiver Operating Characteristics (ROC) curves, of which the horizontal axis is overall average of FA rate and the vertical axis is overall average of FR rate, with classifier baseline and metric learning network with SVM back-end. Although the ROC curves are typically used to evaluate binary classifiers, we extend this to multi-class classification by micro-averaging over all of the classes per model similar to other works. \cite{tang2018deep, choi2019temporal}

\section{Results}
\label{sec:results}
Table \ref{tab:result} shows the overall result of our experiments. Note that the dataset split is different from previous works to reflect the likely scenario in keyword spotting. We set the network trained with cross entropy loss as the baseline. Although the target keyword accuracy is the highest in baseline by a small margin, all of the metric learning based network outperforms the classifier network in terms of non-target accuracy. This indicates that metric learning based keyword spotting system performs well at rejecting unseen-unknown keywords. Instead of using vanilla metric learning methods, training positive pairs only with target-keywords and utilizing support vector machines show significantly better performance on the non-targets. The AP-FC shows highest in terms of total accuracy, non-target accuracy and mAP in our dataset since it is able to take advantages of both classifier and metric learning based training methods.



\begin{figure}[t]
 \centering 
 \includegraphics[width=1\linewidth]{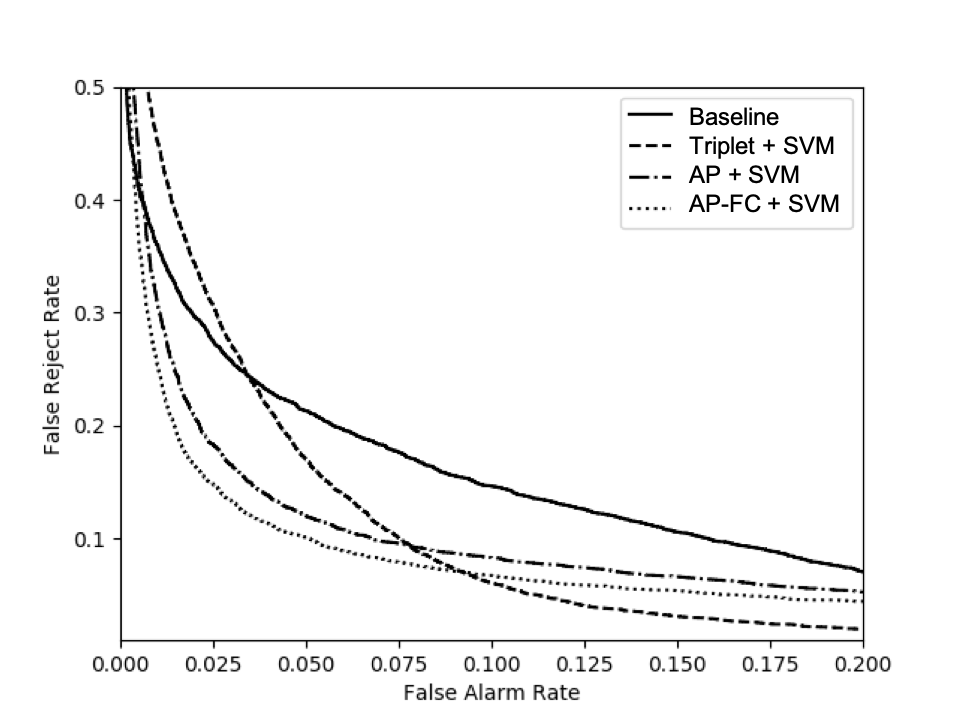}
 \caption{ROC curves of various methods}
 \label{fig:roc}
 \vskip 0.1in
 \end{figure}

Figure \ref{fig:roc} shows ROC curves of the different methods, and the AP-FC outperforms other methods for most useful operating points on the curve.

\section{Conclusion}
\label{sec:conclusion}
In this paper, we have proposed various metric learning approaches in keyword spotting task that significantly improves the detection rate of unseen non-target spoken terms while maintaining target keyword accuracies. We show that utilising metric learning-based methods without gathering the embeddings of non-target sounds works well for detecting unseen non-target sounds. We also propose AP-FC can learn fixed per-class weights in a prototypical framework in order to boost the 
overall accuracy. Our proposed methods outperform baselines on the split of Google Speech Command dataset which better reflects the real-world scenarios. 

\clearpage
\bibliographystyle{IEEEtran}
\bibliography{shortstrings,mybib}

\end{document}